# Chloroplast microsatellites reveal colonisation and metapopulation dynamics in the Canary Island pine

MIGUEL NAVASCUÉS[1], ZAFEIRO VAXEVANIDOU[2,3], SANTIAGO C. GONZÁLEZ-MARTÍNEZ[2], JOSÉ CLIMENT[2], LUIS GIL[3] and BRENT C. EMERSON[1*]

1 Centre for Ecology, Evolution and Conservation, School of Biological Sciences, University of East Anglia, Norwich NR4 7TJ, UK
2 Departamento de Sistemas y Recursos Forestales, CIFOR-INIA, P.O. Box 8111, 28080 Madrid, Spain
3 U.D. Anatomía, Fisiología y Genética Forestal, ETSI de Montes (UPM) Ciudad Universitaria, 28040 Madrid, Spain
* Author for correspondence (b.emerson@uea.ac.uk)

CORRESPONDING AUTHOR:
Brent C. Emerson
School of Biological Sciences
University of East Anglia
Norwich NR4 7TJ, UK
e-mail: b.emerson@uea.ac.uk
fax: (44) 01603 592250

Running title: *P. canariensis* colonisation and population expansion

**ABSTRACT**
Chloroplast microsatellites are becoming increasingly popular markers for population genetic studies in plants, but there has been little focus on their potential for demographic inference. In this work the utility of chloroplast microsatellites for the study of population expansions was explored. First, we investigated the power of mismatch distribution analysis and the $F_S$ test with coalescent simulations of different demographic scenarios. We then applied those methods to empirical data obtained for the Canary Island pine (*Pinus canariensis*). The results of the simulations showed that chloroplast microsatellites are sensitive to sudden population growth. The power of the $F_S$ test and accuracy of demographic parameter estimates, such as the time of expansion, were reduced proportionally to the level of homoplasy within the data. The analysis of Canary Island pine chloroplast microsatellite data indicated population expansions for almost all sample localities. Demographic expansions at the island level can be explained by the colonisation of the archipelago by the pine, while population expansions of different ages in different localities within an island appear to be the result of local extinctions and recolonisation dynamics. Comparable mitochondrial DNA sequence data from a parasite of *P. canariensis*, the weevil *Brachyderes rugatus*, supports this scenario, suggesting a key role for volcanism in the evolution of pine forest communities in the Canary Islands.

## INTRODUCTION

In plants the chloroplast genome is used extensively for evolutionary genetic studies within species in the same way the mitochondrial genome is used within animal studies. However, finding enough sequence variation is a challenge due to the low mutation rates that characterize the chloroplast genome. In contrast, chloroplast microsatellites, or simple sequence repeats (cpSSRs), present higher levels of polymorphism and are easily genotyped and this has made them useful and popular markers for population genetic studies (Provan *et al.*, 2001). Although used extensively for studying population structure and gene flow, the potential of cpSSRs to study population demographic history has received little attention. In this study we investigate in the utility of cpSSR data for the detection of population expansions.

The study of historical demography by means of genetic information is based on coalescent theory (see Emerson *et al.*, 2001for a review). In a stable population coalescence events are scarcer towards the past giving a genealogy dominated by an ancient bifurcation with mutations mainly distributed in inter-node branches (King *et al.*, 2000; Reich & Goldstein, 1998). Contrastingly, in the case of sudden population growth, coalescent events occur mainly during the expansion, leaving a "comblike" genealogy; and mutations are more abundant along the terminal branches (singleton mutations) than in inter-node branches (figure 1 shows the main differences of the two opposing scenarios). As a consequence, population expansions can be detected because of an excess of singletons (Fu & Li, 1993; Tajima, 1989) or an excess of haplotypes (as a consequence of the excess of singletons, Fu, 1997). Also, the divergence between most lineages dates from the time of expansion, producing unimodal distributions of pairwise genetic distances (Slatkin & Hudson, 1991). The study of such distributions also allows for the estimation of the time and magnitude of the population increase (Rogers, 1995; Schneider & Excoffier, 1999).

The methods for studying population expansions are fairly robust for a genetic marker evolving under the unrealistic infinite sites model, where singletons and genetic distances are identified without error. However, in the evolution of sequences under a finite sites model, parallel and back mutations (i.e. homoplasic mutations) will erase part of the genetic information producing inaccurate estimates of singletons and genetic distances. This affects the power of the statistical tests and the estimates of time and magnitude of the demographic growth (Aris-Brosou & Excoffier, 1996; Bertorelle & Slatkin, 1995). In nucleotide sequence data, the usual markers for studying population expansions, the effect of homoplasy is small (Rogers *et al.*, 1996) and can be accounted for in more sophisticated analyses (Schneider & Excoffier, 1999). In cpSSRs, which evolve in a stepwise fashion, higher levels of homoplasy are expected in comparison with sequence data and therefore statistical analyses developed for DNA sequence data may prove unreliable.

In the present work we have simulated the evolution of cpSSRs under constant population size and under population expansion to test the usefulness of these markers for the study of demographic expansions. These theoretical results were then compared with empirical results from the Canary Island pine (*Pinus canariensis*). The presence of *P. canariensis* on each of the five volcanic islands on which it occurs must be through colonisation after the emergence of each island, followed by population expansion.

## MATERIALS AND METHODS

### *Simulations*

Demographic histories of population expansions (recent and old) and stable population size were modelled with coalescent simulations to obtain theoretical expectations of the behaviour of cpSSRs. The coalescent simulation (described in Navascués & Emerson, 2005) consists of the generation of a genealogy for a sample of individuals under a particular demographic history followed by the distribution of mutations randomly onto those lineages. For the population expansions the demographic history was modelled with a logistic equation setting the initial population size ($N_0$) as one individual (coloniser) at the time of expansion ($\tau$, in mutational units). Microsatellite evolution was simulated following a symmetrical single-step mutation model where mutation rates were either heterogeneous (two-rates model) or uniform (one-rate model) across loci. Heterogeneous mutation rates can be considered a more realistic scenario taking into account the differences in polymorphism among cpSSR loci (see, for example, Gómez *et al.*, 2003). As well as being more realistic, heterogeneous mutation rates will also produce higher levels of homoplasy by concentrating the mutations onto particular loci, thus providing a more rigorous assessment of the demographic utility of cpSSRs. The three different demographic histories and the two mutation models gave a combination of six different cases considered (table 1). Simulations were performed for a sample size of 24 individuals and six cpSSR loci. For each case, 1000 replicates were run and their output (genetic state of a sample of individuals in the present generation) was analysed as described in section (c). For each simulated case the level of homoplasy was quantified as the probability that two haplotypes identical in state are not identical by descent (homoplasy index, Estoup *et al.*, 2002).

### *Plant Material and Molecular Markers*

Empirical data was obtained from two previous studies of *P. canariensis* (Gómez *et al.*, 2003; Vaxevanidou *et al.*, 2005). Additionally, three populations from Tenerife (nine, 12 and 13 in table 2 and figure 2) were also genotyped for the present analysis and the compatibility of the data was assured by repeated genotyping of four haplotypes from the previous studies. All individuals were genotyped for six cpSSR loci: Pt15169, Pt30204, Pt71936, Pt87268, Pt26081 and Pt36480 (Vendramin *et al.*, 1996).

### *Data Analysis*

In order to use Arlequin 2.0 (Schneider *et al.*, 1999) for the analyses, microsatellite data was binary coded: the number of repeats were coded with "1" and shorter alleles were coded filling the difference in repeats with "0" (Pereira *et al.*, 2002). Analyses for the empirical samples were carried out at two levels: (1) sample sites as the unit of analysis, (2) islands as the unit of analysis with sample sites within an island pooled together.

A general description of diversity indices and population structure found within *P. canariensis* using cpSSRs is presented in Gómez *et al.* (2003); thus here we focus on the assessment of demographic history, using two different but complementary approaches. Firstly, we performed the $F_S$ neutrality test for population expansion (Fu, 1997). This test is based on different expectations for the number of haplotypes when comparing a stationary with an expansion demography. The $F_S$ statistic takes a large negative value within a population affected by expansion due to an excess of rare haplotypes (recent mutations). Significance of the test was calculated with 10 000 data bootstraps (Schneider *et al.*, 1999). An $F_S$ statistic with $p(F_S) < 0.02$ ($\alpha = 0.05$, due to a particular behaviour of this statistic, Fu, 1997) was considered evidence of population expansion.

The second analysis consists of the estimation of the demographic model of Rogers & Harpending (1992) described with the parameters: $\tau = 2\mu t$, $\theta_0 = 2\mu N_0$ and $\theta_1 = 2\mu N_1$ (where $\mu$ is the mutation rate, $t$ is the number of generations since expansion and $N_0$ and $N_1$ are the population sizes before and after expansion). Parameters are estimated from the distribution of pairwise differences (difference in number of repeats) between individuals within a sample. Although, in our case, the pairwise differences calculated cannot be strictly called mismatches, we will refer to their distribution as a mismatch distribution as it is the most common term used throughout the literature (Harpending *et al.*, 1993). This distribution is affected by the demography of the sample; sudden growth produces unimodal distributions while within stationary populations distributions are ragged and multimodal (Slatkin & Hudson, 1991). An algorithm, which minimizes the sum of squared differences (SSD) between model and data, estimates the combination of parameters with the best fit to the empirical data (Schneider & Excoffier, 1999). The strength of the estimated model is then evaluated from the SSD distribution which is obtained from 10 000 data bootstraps (1000 for the simulation output), making $p$(SSD) the proportion of bootstraps with the SSD larger than the original (Schneider & Excoffier, 1999). A significant SSD value, $p$(SSD) < 0.05, implies the rejection of the estimated demographic model. The confidence interval (95% CI) for the estimated parameter $\hat{\tau}$ is also calculated from the bootstrap process (Schneider & Excoffier, 1999). Confidence intervals for parameters related to the magnitude of expansion ($\hat{\theta}_0$ and $\hat{\theta}_1$) will not be discussed as they are usually too wide and are of less interest for the interpretation of the results (Excoffier & Schneider, 1999). Dating the population expansions was done using the parameter $\hat{\tau}$ and its relationship with time and mutation rate: $\tau = 2l\mu t$ (where $l$ is the number of cpSSR loci and $\mu$ is the mutation rate per locus).

## RESULTS AND DISCUSSION

### *Simulations*

The results from the simulations are summarised in table 1. In the two analyses performed, cpSSR polymorphism was sensitive to population growth; however the results were not as precise as would be desirable.

#### *$F_S$ Neutrality Test*

In the cases of uniform mutation rate across loci the performance of the $F_S$ test to detect population expansion was acceptable. Type II error for the $F_S$ test (no evidence of population expansion in cases 1 and 2) was very low, and type I error (rejection of stationary population size in case 3) was low (11% of the replicates of case 3), although greater than expected at the given confidence level (expected 5% for $\alpha$=0.05).

In the cases evolving under the two-rate model (cases 4–6), the power of the $F_S$ test decreased dramatically, and this was accompanied by an increase in homoplasy. Detection of recent expansions was especially affected and the reason for this relates to the estimates of genetic distance and the number of haplotypes used in the test. First, the test uses the average genetic distance among individuals to calculate the expected number of haplotypes under a stationary demography scenario. The effect of homoplasy in this calculation is proportional to the time of expansion, with an average reduction of 19% in the distance estimates of recent expansions (case 4) and 41% in the older expansions (case 5). The expected number of haplotypes is then compared to the observed number of haplotypes. While the effect of homoplasy in genetic distance estimates was proportional to the time of expansion, homoplasy decreases the detectable number of haplotypes by approximately 40% both in the recent and older expansions. It seems that the power of the test varies with the time of expansions because the error in the estimates of genetic distances and number of haplotypes is more unbalanced for recent expansions.

*Demographic Model Estimation*

For cpSSRs evolving under the one-rate mutation model, estimates of the time of expansion were fairly accurate, although older expansion times were slightly underestimated. The average estimated time of expansion ($\hat{\tau}$) for the recent expansions (case 1, $\tau = 1.0$) was 1.1 and the true value was always within the 95% CI, while for older expansions (case 2, $\tau = 3.0$) average $\hat{\tau}$ was 2.5 and the true value falls outside the 95% CI in 15% of the replicates. In the simulations using the two-rate mutation model the estimates for recent expansion (case 5, $\tau = 1.0$) were accurate, with average $\hat{\tau} = 1.0$ and the true value fell outside the 95% CI in only 2% of the replicates. However in older expansions (case 6, $\tau = 3.0$) the expansion time was largely underestimated for the two-rate mutation model with the average value of $\hat{\tau}$ being 1.8 and the true value falling outside the 95% CI in 77% of the replicates. Although these results appear discouraging it is important to note that the relative times of expansion are still discernable, and that it may be possible to develop new statistical analyses to improve the estimates as has been done for heterogeneous mutation rates within sequence data (Schneider & Excoffier, 1999).

***The Empirical Case: Pinus canariensis***

The results for the detection of population expansions in the *P. canariensis* samples are reported in table 2. For the estimation of the demographic model the algorithm was unable to find a combination of parameters with a minimum SSD in three samples (Tamadaba, Chinyero and El Hierro). This inability of the algorithm to converge sometimes has been observed in previous works (e.g. Stamatis *et al.*, 2004) and in our simulations. A simple solution is to obtain the estimation from a reduced sample obtained by randomly removing one individual. This reduction of the sample size changes the shape of the mismatch distribution slightly enough for the algorithm to converge while still maintaining a very similar shape to the mismatch distribution from the original data set. The mismatch distributions from the reduced samples were used to produce the parameter estimations presented in italics in table 2.

The demographic expansion model estimated for different sampling sites (including the grouping of sampling sites at the island level) was, in general, fairly robust [$p$(SSD) >> 0.05] and mismatch distributions were clearly unimodal (figure 2;

opposite to the ragged distribution expected with a stable population). The results of the $F_S$ test yielded evidence of population expansion for nearly half of the samples. It is interesting to note that the samples for which the $F_S$ test could not reject a stable population scenario [$p(F_S) > 0.02$] were the ones with the lowest $\hat{\tau}$ values. In the light of our simulation results it is expected that the $F_S$ test will have lower power to detect very recent population expansions, especially under the more realistic scenario of heterogeneous mutation rates across loci. Thus we could consider that most of the *P. canariensis* populations are likely to have been subject to demographic growth and the lack of statistical evidence is due to the low power of the $F_S$ test for the most recent expansions.

*Island Level: Colonisation*

Compared to continental areas, oceanic island populations are typically established by only one or a few individual founders that successfully reproduce, leading to demographic expansions. Whether the population expansions detected for *P. canariensis* at the island level reflect the initial colonisation of the islands or subsequent demographic events is difficult to know. However, times of expansion in relation to the geological history of the archipelago can supply the necessary clues to discern between both possibilities.

Potential maximum times for expansion are bound to the emergence times of the islands. The maximum subaerial geological age of El Hierro, the youngest island, is approximately one million years (Carracedo & Day, 2002). If we consider that the time of the population expansion in El Hierro is $\hat{\tau} = 1.291$ and the relationship $\tau = 2l\mu t$ we obtain a mutation rate estimate of $1.076 \times 10^{-5}$ per locus per generation (considering generation time to be 100 years as in Provan *et al.*, 1999). Using this mutation rate estimate we calculated the maximum age of population expansion for each sample, reported in table 2.

In order to establish a minimum time of expansion we have analysed mtDNA COII sequence data for *Brachyderes rugatus* from Emerson *et al.* (2000 and unpublished data). Because the niche of this species is the pine tree, its demographic expansions must have occurred either during or after the establishment of the pine forest on each island. Population expansions have been detected (significant $F_S$ test) for the islands of La Palma and Tenerife (138 and 182 individuals respectively, sampled throughout the islands). The times of expansion for *B. rugatus* were estimated from the mismatch distributions to be approximately 0.72 million years ago (mya) for Tenerife and 1.11 mya for La Palma (considering divergence rates to be between 2% and 2.3% per million years, Brower, 1994; DeSalle *et al.*, 1987). These dates strengthen the age estimates for the expansion of the pine forest obtained with the geological age calibration.

These age estimates suggest expansions of the pine tree increasing in age from West to East, and coinciding broadly with the colonisation ages estimated for *B. rugatus* (Emerson *et al.*, 2000), as shown in table 2. We interpret the expansions at the island level as a result of the colonisation process and linked to the volcanic history of the archipelago. The creation of new emerged landmass by recent (up to 2 mya) volcanic activity in the younger islands (La Palma and El Hierro) opened new territories for *P. canariensis* to colonise. Note that the age of Tenerife presented in figure 2 refers to its older massifs which are the remains of two or three smaller precursor islands.

However, the majority of the landmass of Tenerife was mainly formed by the activity of Las Cañadas volcano starting around 2 million years ago (Ancochea *et al.*, 1990) and it is this event which would appear to be causally related to the pine forest expansion. On the island of Gran Canaria, an episode of heavy volcanic activity (Roque Nublo volcano, Pérez-Torrado *et al.*, 1995) is believed to have destroyed almost all terrestrial ecosystems within the island, with perhaps the exclusion of some coastal regions, between 5.5 and 3 million years ago (Marrero & Francisco-Ortega, 2001), and this hypothesis has gained recent support from a meta-analysis by Emerson (2003). The expansion of the pine forest in Gran Canaria after that event can be explained either by colonisation of *P. canariensis* to the island or by a bottleneck if a small pocket of pine forest survived through the Roque Nublo eruptive period.

*Sample Level: Metapopulation Dynamics*

The islands of the Canary archipelago have a geological history marked by recent dramatic volcanic activity and giant landslides (Carracedo & Day, 2002). These destructive events would have produced local elimination of pine forest, as has been recorded for historical volcanic eruptions (del Arco Aguilar *et al.*, 1992; Pérez de Paz *et al.*, 1994). Also, the Canary Island pine is renowned by its capacity for colonising lava flows (del Arco Aguilar *et al.*, 1992; Pérez de Paz *et al.*, 1994), which suggest that a metapopulation dynamic occurs within the pine forest. One of the genetic signals expected in the local recolonisations after volcanic disturbances are those of the demographic expansions, as it has been shown in other organisms subject to similar metapopulation dynamics in other volcanic archipelagos (Beheregaray *et al.*, 2003; Vandergast *et al.*, 2004).

It seems very likely that local expansions detected for *P. canariensis* are the product of metapopulation dynamics. When we consider different samples within the same island (in Tenerife and Gran Canaria) we observe that the expansion of pine forest at some areas is younger than the main demographic expansion affecting the island. We hypothesize that the apparently more recent expansions may be areas recolonised after geological disturbance. The role of volcanism and giant landslides in the reduction of genetic diversity has also been proposed to explain the pattern of diversification of *Brachyderes rugatus* in La Palma, El Hierro, Tenerife and Gran Canaria (Emerson *et al.*, 2000).

## CONCLUSIONS

This study demonstrates the utility of cpSSRs for the detection of demographic expansions and the estimation of their relative ages. The application of population genetic demographic methodology to cpSSR data for *P. canariensis* populations revealed new insights into the population history of this species. The volcanic activity of the archipelago appears to be a disturbance agent in the pine forest ecosystem, conditioning the areas available for the pine tree. Future studies of mitochondrial DNA data may further complement data from cpSSRs to elucidate the colonisation and population dynamic history of *P. canariensis* on the Canary Islands. A mtDNA phylogeographic analysis would reflect the historical seed movements of *P. canariensis*, which are limited relative to pollen and may contain more fine scale phylogeographic information. Additionally, a sampling design including historical and isotope-dated lava flows within the pine forest may provide a good test for the hypothesis of a metapopulation dynamic.

Our analyses have revealed homoplasy as a problem for the analyses (mainly in the detection of younger expansions) because it reduces the power of the $F_S$ test and accuracy of absolute expansion time estimates. The development of statistics taking into account the effects of homoplasy would further improve the usefulness of cpSSRs as well as other linked microsatellite markers such as Y-chromosome microsatellites for demographic studies.

**Acknowledgements**

MN scholarship was funded by the University of East Anglia. We thank the Cabildo Insular de Tenerife for collecting permits.

Figure 1: Coalescent process under two contrasting scenarios: constant population size and sudden population expansion. For each case, the demographic history and, in the same timescale measured in mutational units (1 mutational unit = $1/2\mu$ generations) the simulated genealogy of a random sample of genes is represented, with stars representing mutational events. Below, chloroplast microsatellite mismatch distribution and result of the $F_S$ test and demographic parameters estimates for those simulated samples are shown.

Figure 2: Chloroplast microsatellite mismatch distributions for the islands and map of the Western Canary Islands. Maximum subaerial age of the islands (Carracedo & Day, 2002) are shown in parenthesis (in million of years). Sampling localities are marked with numbers, corresponding to those shown in table 2.

Table 1: Population expansion signal on the $F_S$ test and mismatch distribution analysis and homoplasy level in the six simulated cases.

| Case | Expansion time, $\tau$ | Mutation rate, $\mu$ | | Proportion of non-significant $F_S$ test | Proportion of significant SSD | Homoplasy index, $P$ |
|---|---|---|---|---|---|---|
| | | loci 1-2 | loci 3-6 | | | |
| 1 | 1 (recent) | $5.5 \times 10^{-5}$ | | 0.038 | 0.052 | 0.049 |
| 2 | 3 (old) | $5.5 \times 10^{-5}$ | | 0.002 | 0.051 | 0.297 |
| 3 | no expansion | $5.5 \times 10^{-5}$ | | 0.893 | 0.144 | 0.065 |
| 4 | 1 (recent) | $1.65 \times 10^{-4}$ | $10^{-7}$ | 0.553 | 0.095 | 0.122 |
| 5 | 3 (old) | $1.65 \times 10^{-4}$ | $10^{-7}$ | 0.240 | 0.060 | 0.606 |
| 6 | no expansion | $1.65 \times 10^{-4}$ | $10^{-7}$ | 0.931 | 0.080 | 0.263 |

Table 2: Results for the $F_S$ neutrality test for population expansion (Fu, 1997) and population expansion parameters $\hat{\tau}$ following Schneider & Excoffier (1999). Estimates are presented in italics when the algorithm did not converge (see Results and discussion section for details). The time of expansion expressed in million of years before present (mya) is calculated using mutation rates in the range $1.076 \times 10^{-5}$ per generation per locus. The results for the islands (pooling samples from the same island) are presented in bold. Garabato is a monomorphic population and tests could not be performed. For comparison, time for the colonisation of *Brachyderes rugatus* are also presented from Emerson *et al.* (2000).

| Population | *N* | Fu (1997) | | Schneider & Excoffier (1999) | | | | Emerson *et al.* (2000) |
|---|---|---|---|---|---|---|---|---|
| | | $F_S$ | $p(F_S)$ | $\hat{\tau}$ (95% CI) | *t* (mya) | SSD | *p*(SSD) | *B. rugatus* (MYA) |
| **Gran Canaria** | **145** | **-26.260** | *** < 0.001** | **2.544 (1.372-5.010)** | **1.970** | **0.002** | **0.569** | > 2.56 |
| 1 Arguineguín | 30 | -21.890 | * < 0.001 | 3.703 (2.138-5.636) | 2.868 | 0.003 | 0.469 | |
| 2 Galdar | 19 | -4.076 | * 0.017 | 2.722 (1.128-3.892) | 2.108 | 0.004 | 0.554 | |
| 3 Mogán | 24 | -0.849 | 0.275 | 0.969 (0.000-1.558) | 0.750 | 0.020 | 0.077 | |
| 4 Tamadaba | 24 *(23)* | -1.304 | 0.142 | *0.871 (0.000-1.489)* | *0.675* | *0.014* | *0.138* | |
| 5 Tirajana | 24 | -5.052 | * 0.014 | 1.743 (0.413-5.224) | 1.350 | 0.008 | 0.399 | |
| 6 Tirma | 24 | -6.845 | * 0.001 | 2.438 (0.779-3.342) | 1.888 | 0.002 | 0.783 | |
| **Tenerife** | **280** | **-26.710** | *** < 0.001** | **2.374 (1.330-3.144)** | **1.839** | **0.000** | **0.910** | 1.89-2.56 |
| 7 Anaga | 24 | -0.192 | 0.185 | 3.000 (0.523-3.000) | 2.323 | 0.010 | 0.062 | |
| 8 Arico | 24 | -8.983 | * < 0.001 | 2.314 (0.675-3.185) | 1.792 | 0.003 | 0.542 | |
| 9 Chinyero | 50 *(49)* | -15.290 | * < 0.001 | *2.294 (0.978-2.900)* | *1.777* | *0.001* | *0.636* | |
| 10 La Esperanza | 24 | -3.040 | 0.022 | 2.722 (0.483-6.413) | 2.108 | 0.015 | 0.567 | |
| 11 La Guancha | 24 | -3.545 | 0.024 | 2.036 (0.501-2.875) | 1.577 | 0.003 | 0.623 | |
| 12 Güímar | 47 | -22.280 | * < 0.001 | 3.280 (1.795-4.143) | 2.540 | 0.001 | 0.702 | |
| 13 Ifonche | 39 | -8.615 | * < 0.001 | 2.482 (1.035-3.172) | 1.922 | 0.007 | 0.130 | |
| 14 Oratava | 24 | -6.456 | * 0.002 | 2.730 (0.993-6.926) | 2.114 | 0.010 | 0.285 | |
| 15 Vilaflor | 24 | -0.165 | 0.437 | 1.081 (0.000-1.766) | 0.837 | 0.009 | 0.192 | |
| **La Gomera** | **36** | **-0.424** | **0.427** | **1.558 (0.380-3.379)** | **1.207** | **0.008** | **0.431** | |
| 16 Garabato | 12 | - | - | - | - | - | - | |
| 17 Imada | 24 | -0.879 | 0.294 | 1.307 (0.182-2.064) | 1.012 | 0.002 | 0.711 | |
| **La Palma** | **48** | **-2.826** | **0.072** | **1.244 (0.311-1.726)** | **0.963** | **0.004** | **0.285** | 1.58-2.00 |
| 18 Fuencaliente | 24 | -1.063 | 0.224 | 1.289 (0.048-2.002) | 0.998 | 0.008 | 0.305 | |

| | | | | | | | | | |
|---|---|---|---|---|---|---|---|---|---|
| 19 Garafía | 24 | -1.279 | 0.184 | 1.206 (0.053-1.904) | 0.934 | 0.002 | 0.720 | | |
| **20 El Hierro** | **24** ***(23)*** | **-3.513** | *** 0.008** | ***1.291 (0.040-2.035)*** | ***1.000*** | ***0.008*** | ***0.301*** | | 1.00 |

* Significant at $\alpha = 0.05$ ($p$-value < 0.02)

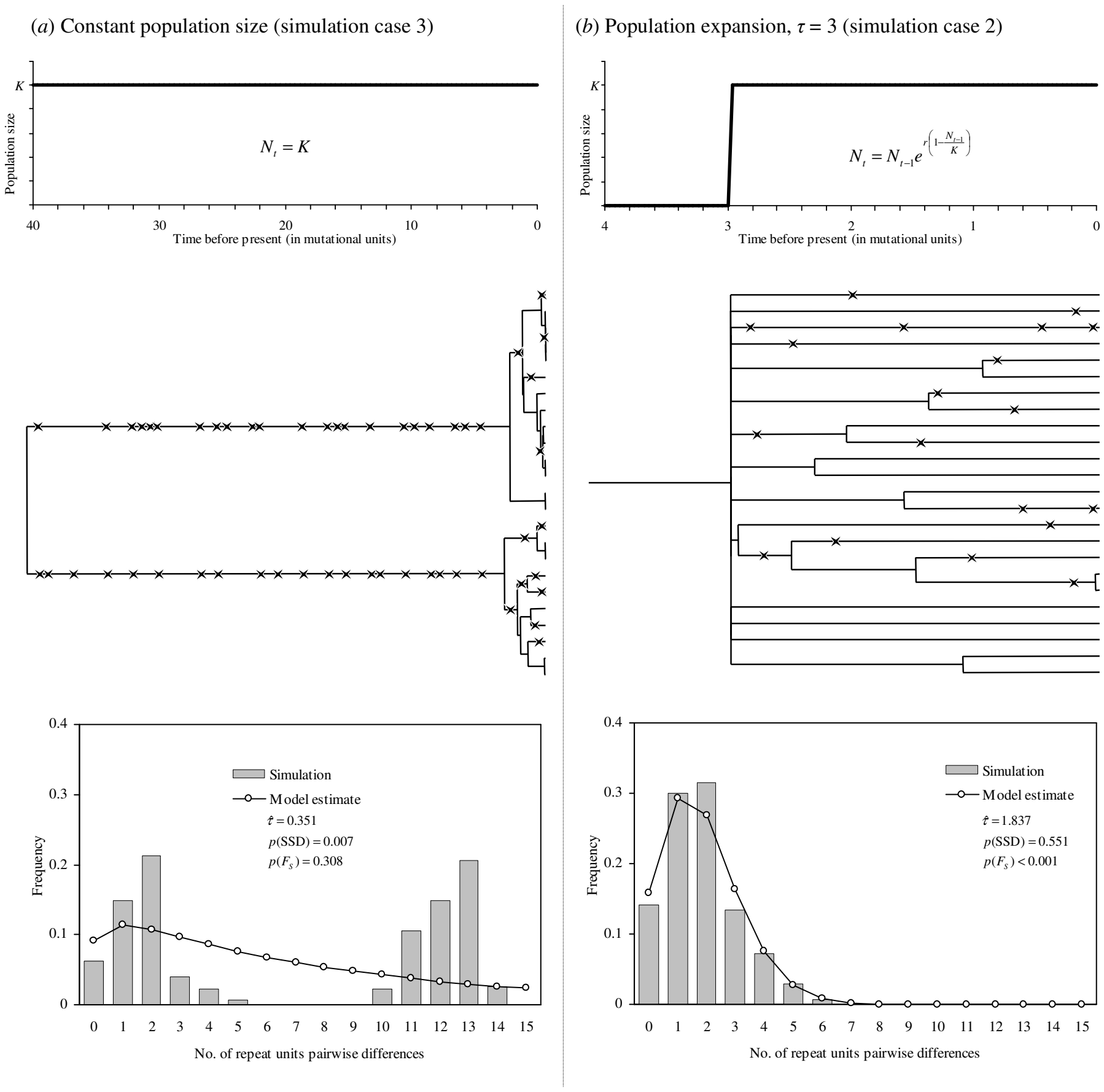

(a) Constant population size (simulation case 3)
(b) Population expansion, τ = 3 (simulation case 2)
K
Population size
$N_t = K$
$N_t = N_{t-1}e^{r\left(1-\frac{N_{t-1}}{K}\right)}$
40
30
20
10
0
4
3
2
1
Time before present (in mutational units)
0.4
0.3
0.2
0.1
Frequency
Simulation
Model estimate
$\hat{\tau} = 0.351$
$p(\mathrm{SSD}) = 0.007$
$p(F_S) = 0.308$
$\hat{\tau} = 1.837$
$p(\mathrm{SSD}) = 0.551$
$p(F_S) < 0.001$
0 1 2 3 4 5 6 7 8 9 10 11 12 13 14 15
No. of repeat units pairwise differences

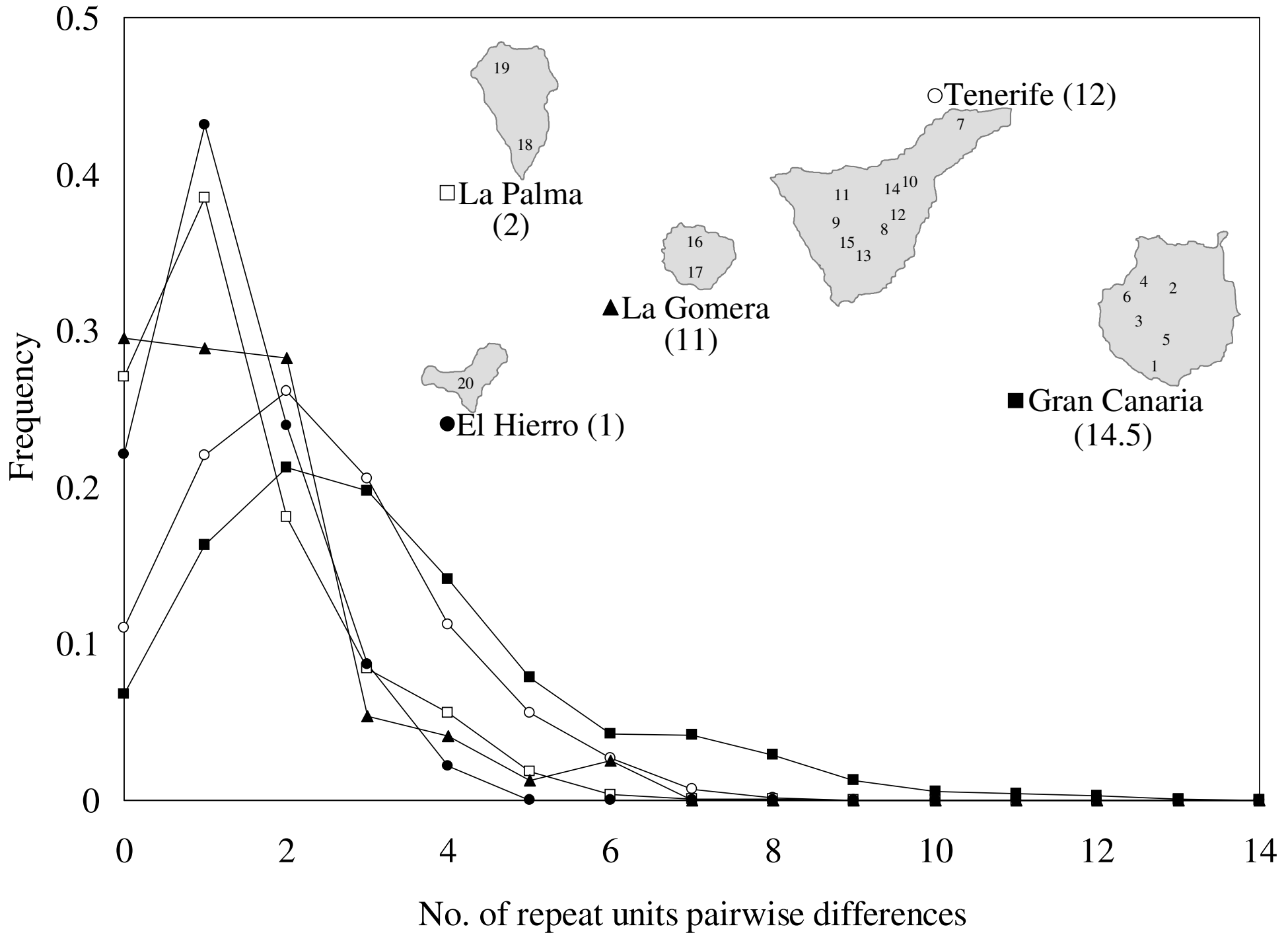